\documentclass[twocolumn,showpacs,amssymb,preprintnumbers,prc,floatfix,nofootinbib]{revtex4-1}
\usepackage{epsfig,dcolumn}
\usepackage{bm,color}
\usepackage{graphicx,bm}
\usepackage{gensymb}
\newcommand{\be}{\begin{equation}}
\newcommand{\ee}{\end{equation}}
\newcommand{\ba}{\begin{eqnarray*}}
\newcommand{\ea}{\end{eqnarray*}}

\begin{document}
\title{Limits on assigning a shape to a nucleus}

\author{A.~Poves$^1$,  F.~Nowacki$^{2}$, and Y. Alhassid$^{3}$}

\affiliation{ $^1$Departamento de F\'isica Te\'orica and IFT-UAM/CSIC, Universidad Aut\'onoma de Madrid,  E-28049
	     Madrid, Spain\\$^2$Universit\'e de Strasbourg, IPHC, 23 rue du Loess and  CNRS, UMR7178, 67037 Strasbourg, France\\
	$^{3}$Center for Theoretical Physics, Sloane Physics Laboratory, Yale University, New Haven, Connecticut 06520, USA}

\begin{abstract}
The interpretation of nuclear observables in the laboratory frame in terms of the intrinsic deformation parameters $\beta$ and $\gamma$
is a classical theme in nuclear structure. Here we use the
  quadrupole invariants (Kumar~\cite{kumar72}), calculated within the framework of the configuration-interaction shell model, to clarify the meaning and limitations of nuclear shapes.  We introduce a novel method that enables us to calculate accurately higher-order invariants and, therefore, the fluctuations in both $\beta$ and $\gamma$. We find that the shape parameter $\beta$ often has a non-negligible degree of softness, and that the angle $\gamma$ is usually characterized by large fluctuations, rendering its effective value not meaningful. Contrary to common belief, we conclude that doubly magic nuclei are not spherical because the notion of a well-defined shape does not apply to them.

\end{abstract}

\date{\today}
\maketitle 

\section{ Introduction} The description of collectivity in nuclear structure has its roots in the unified model of Bohr and Mottelson~\cite{bm53}. The treatment of the dominant quadrupole correlations and the description of nuclear deformation have often been carried out in the intrinsic frame, and 
quadrupole shapes have been characterized by the intrinsic deformation parameters $\beta$ and $\gamma$. We commonly characterize a nucleus as 
prolate when $\gamma$=0$^\circ$, oblate when $\gamma$=60$^\circ$, and rigid triaxial when $\gamma$=30$^\circ$.
There has been a longstanding debate as to whether nuclei with rigid triaxial deformation exist, or instead can only be $\gamma$-soft. Often the same nucleus exhibits signatures of intrinsic structures with different values of $\beta$ and $\gamma$, a phenomenon known as shape coexistence~\cite{hw}. The intrinsic deformation parameters can change along isotopic or isotonic chains, referred to as shape evolution or shape transition.

The intrinsic shape parameters are usually inferred from experimental values of observables such as excitation energies, $E2$ transitions, and spectroscopic quadrupole and magnetic moments.  Intrinsic shapes are commonly calculated in the framework of a mean-field approximation. However, such an approximation breaks rotational invariance and ignores important correlations.

  Shape parameters  can also be extracted from calculations carried out in the laboratory frame, in which case it is necessary to agree on
  a set of rules for transforming between the laboratory frame and the intrinsic frame. Such rules were outlined by Davidov and Filipov~\cite{dafi}
  and many others.  However, the only rigorous method to relate the intrinsic parameters to laboratory-frame observables is provided
  by the so-called quadrupole invariants $\hat Q^n$ of the second-rank quadrupole operator $\hat Q$, introduced by  Kumar~\cite{kumar72} (see also Ref.~\cite{cline86}), whose expectation values are independent of the particular frame. The effective values of $\beta$ and $ \gamma$  are determined from the expectation values of the second- and third-order invariants defined, respectively, by $\hat Q^2=\hat Q\cdot \hat Q$ and $\hat Q^3=  (\hat Q\times \hat Q)\cdot \hat Q$ (where $\hat Q \times \hat Q$ is the coupling of $\hat Q$ with itself to a second-rank operator).  These invariants were recently applied to describe the evolution of collectivity in cadmium isotopes~\cite{heyde-cd}.  
  
 However, it is not meaningful to assign effective values to $\beta$ and $\gamma$ without also studying their fluctuations.  Fluctuations in $\beta$ can be determined from the variance $\sigma^2$ of $\hat Q^2$ 
   \begin{equation}\label{sigmaQ2}
  \sigma{(\hat Q^2)} = (\langle \hat Q^4 \rangle - \langle \hat Q^2 \rangle^2)^{1/2}  \;,
  \end{equation}
 and requires the evaluation of the the expectation value of the fourth-order invariant $\hat Q^4$.  Such fluctuations in $\beta$ were studied for example in Refs.~\cite{yoram:14,yoram:17,kasia:16,kasia:18}.
  However, calculation of the fluctuations in $\gamma$ requires the variance of $\hat Q^3$
  \begin{equation}\label{sigmaQ3}
  \sigma{(\hat Q^3)} = (\langle \hat Q^6 \rangle - \langle \hat Q^3 \rangle^2)^{1/2}  \;,
  \end{equation}
  as well as the covariance of $\hat Q^2$ and $\hat Q^3$ (see Eq.~(\ref{sigma-cos3g}) below).  Thus, the expectation values of fifth- and sixth-order invariants are needed to evaluate the fluctuations in $\gamma$.  The calculation of such higher-order invariants has been a major challenge. Here we introduce a novel method to calculate accurately higher-order invariants in the framework of the configuration-interaction (CI) shell model.  This enables us to study systematically the fluctuations of both shape parameters $\beta$ and $\gamma$, and to fully address the question of whether nuclear shapes are well defined.  We apply our method to selected nuclei in the mass region $A \sim 20 -76$.  In many of these nuclei, we find a non-negligible degree of softness in $\beta$, while fluctuations in $\gamma$ are almost invariably large.  In contrast to recent claims~\cite{ge76,ge76a,se76}, we show that the $A=76$ isobars of germanium and selenium are not rigid triaxial. We also demonstrate that doubly magic nuclei, commonly considered spherical,  do not have any particular shape.

\section{Higher-order invariants}
 The choice of  the fourth-order invariant $\hat Q^4$ is unique~\cite{commute} and 
we define it as $\hat Q^4= (\hat Q^2)^2=(\hat Q \cdot \hat Q)^2$. 
The fifth-order invariant is also unique and we take it as  \mbox{$\hat Q^5 = \hat Q^2 \; \hat Q^3= (\hat Q \cdot \hat Q) ([\hat Q \times \hat Q] \cdot \hat Q])$}. The sixth-order invariant is not unique. There are two choices but the adequate one is  $\hat Q^ 6=  (\hat Q^3)^2=([\hat Q \times\hat  Q] \cdot \hat Q])^2$.
 
 To compute the expectation values of these invariants, we take advantage of the fact that our shell model codes incorporate naturally
 the projected Lanczos strength function method~\cite{rmp}. To make our analysis as simple as possible, we confine our study 
 to the ground states of even-even nuclei and to certain excited 0$^+$ states which are of particular interest. Our method follows several steps:
 
\noindent (i) We perform a shell model calculation to obtain the wave function of the  $| 0^+ \rangle$  state of interest.
 
\noindent   (ii) We apply the axial quadrupole operator $\hat Q_{20}$ to this state to obtain
 \begin{equation}\label{SR1}
\hat Q_{20} | 0^+ \rangle = {\rm SR}1(2^+)^{1/2} | \overline{2^+(1)} \rangle \;,
 \end{equation}
where the constant ${\rm SR1}(2^+)$ is defined such that the state $| \overline{2^+(1)} \rangle$ is normalized. This state is not an eigenstate of the Hamiltonian but can be considered a 'doorway'  or a 'sum rule' (${\rm SR}$) state.

\noindent  (iii) Using (\ref{SR1})  we have 
  \begin{equation}\label{Q2} 
\langle \hat Q^2 \rangle = 5 \, {\rm SR}1(2^+) \;.
 \end{equation}

\noindent (iv) We compute the reduced matrix element of $\hat Q$ in the doorway state  $| \overline{2^+(1)} \rangle$ and define
 \begin{equation}\label{reduced-Q}
 \langle \langle \hat Q \rangle \rangle = \frac{1}{\sqrt{5}} \langle \overline{2^+(1)} || \hat Q || \overline{2^+(1)}  \rangle  \;.
 \end{equation}

\noindent  (v) Using $\langle \hat Q^3\rangle= -5\sqrt{7/2} \langle \hat Q_{20}^3 \rangle$~\cite{yoram:14} together with (\ref{Q2}) and (\ref{reduced-Q}), we obtain
 \begin{equation}\label{Q3}
\langle \hat Q^3 \rangle =\langle\hat  Q^2 \rangle  \langle \langle \hat Q \rangle \rangle \;.
\end{equation}   
 
 \noindent (vi) Making a second iteration with $\hat Q_{20}$ on the doorway state in (\ref{SR1}), we produce three
   new normalized doorway states, $ | \overline{0^+(2)} \rangle $,  $ | \overline{2^+(2)} \rangle $,  and  $ | \overline{4^+(2)} \rangle $,  along with their corresponding
     projected sum rules   ${\rm SR}2(0^+)$,   ${\rm SR}2(2^+)$,  and  ${\rm SR}2(4^+)$.

\medskip
\noindent (vii) The expectation value of the fourth-order invariant is then calculated from
  \begin{equation}\label{Q4}
\langle \hat Q^4 \rangle = 5\langle \hat Q^2 \rangle \; {\rm SR}2(0^+) \;.
 \end{equation} 
 
\noindent (viii) Iterating again with $\hat Q_{20}$ on the doorway state $ | \overline{2^+(2)} \rangle$, we obtain  new doorway states, $ | \overline{0^+(3)} \rangle $,  $ | \overline{2^+(3)} \rangle $,  and  $ | \overline{4^+(3)} \rangle $,  and their corresponding projected sum rules  ${\rm SR}3(0^+)$,  ${\rm SR}3(2^+)$,  and  ${\rm SR}3(4^+)$.
 
\medskip
\noindent (ix) Using the doorway states in (viii), we calculate the expectation value of the sixth-order invariant from
\begin{equation}\label{Q6}
\langle \hat Q^6\rangle =5 \langle \hat Q^2 \rangle \; {\rm SR}2(2^+) \;{\rm SR}3(0^+) \;.
\end{equation} 
 
\noindent (x)  The expectation value of the fifth-order invariant can also be calculated using
  \begin{eqnarray}\label{Q5}
\langle \hat Q^5\rangle & = &  5 \langle \hat Q^2 \rangle \; {\rm SR}2(0^+)^{1/2 } \;  {\rm SR}2(2^+)^{1/2}  \; {\rm SR}3(0^+)^{1/2} \nonumber \\ 
& &\times |\langle \overline{0^+(2)}  | \overline{0^+(3)} \rangle | \; {\rm sign} \langle \hat Q^3  \rangle \;.
\end{eqnarray}
  
\begin{figure}[]
\begin{center}
\includegraphics[width=0.30\textwidth]{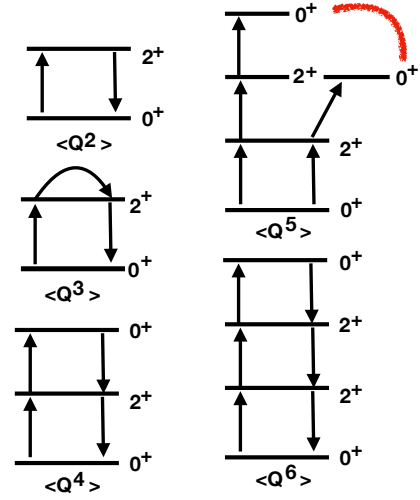}
\end{center}
\caption{\label{qn-diag}  Schematic diagram demonstrating the computation of the quadrupole invariants up to sixth order. Except for the
 initial 0$^+$ states,  all states are sum rule or doorway states, which are not eigenstates of the Hamiltonian but have good spin and parity. Each arrow represents the action of the axial quadrupole $\hat Q_{20}$ operator between the two states which it connects. The red wiggly line denotes the overlap between the two different 0$^+$ doorway states.}
\end{figure}

  In Fig.~\ref{qn-diag} we present a schematic diagram of how we produce all the doorway states that are needed to compute the expectation values of the quadrupole invariants up to sixth order.
 
\noindent (xi) The final expressions for the width-to-average ratios for $\hat Q^2$ and $\hat Q^3$ are given by
  \begin{equation}
  \frac{\sigma{\langle \hat Q^2 \rangle}}{\langle \hat Q^2 \rangle} = \displaystyle \left( \frac{5 \;   {\rm SR}2(0^+)}{\langle Q^2 \rangle} -1 \right)^{1/2} \;,
  \end{equation} 
 and
 \begin{equation}
  \frac{\sigma{\langle \hat Q^3 \rangle}}{\langle \hat Q^3 \rangle} = \displaystyle \left( \frac{5 \;   {\rm SR}2(2^+) \; {\rm SR}3(0^+)}{\langle \hat Q^2 \rangle \; \langle \langle \hat Q \rangle \rangle^2} -1 \right)^{1/2} \;.
  \end{equation} 
    The intrinsic quadrupole moment $Q_0$ and effective values of the Bohr-Mottelson shape parameters $\beta$ and $\gamma$ can be calculated from the expectation values of the second- and third-order invariants using
    \begin{eqnarray}
  Q_0 = \sqrt{\frac{16 \pi}{5}} \langle \hat Q^2 \rangle^{1/2} \;,
    \end{eqnarray}  
     \begin{eqnarray}
  \beta = \frac{4 \pi}{3 r_0^2}  \frac{\langle \hat Q^2 \rangle^{1/2}}{A^{5/3}} \;,
    \end{eqnarray}  
   with r$_0$=1.2~fm, and
      \begin{eqnarray}
  \cos 3\gamma = - \sqrt{\frac{7}{2}}  \frac{\langle \hat Q^3 \rangle}{\langle \hat Q^2 \rangle^{3/2}} =
   - \sqrt{\frac{7}{2}}  \frac{ \langle \langle \hat Q \rangle \rangle}{\langle \hat Q^2 \rangle^{1/2}}   \;.
 \end{eqnarray}  
 To gain insight into how well defined are these effective shape parameters, we also calculate their variances in the framework of the CI shell model. 
 These variances tell us to what extent the intrinsic shapes are blurred in the laboratory frame, and whether and when the very notion of shape makes sense at all.   
     
\section{Fluctuations of $\beta$}

  The standard deviation $\Delta\beta$ of the effective $\beta$ parameter is easily determined from the standard deviation of $\hat Q^2$  
\begin{equation}
      \frac{\Delta \beta}{\beta} = \frac{1}{2} \frac{\sigma \langle \hat Q^2 \rangle}{ \langle \hat Q^2 \rangle}\;.
\end{equation}    

The fluctuations of $\beta$ have been studied in Refs.~\cite{kasia:16,kasia:18} for the ground state and the deformed excited band of $^{42}$Ca 
 using large-scale shell-model calculations, and in Ref.~\cite{yoram:17} for a series of  neodymium and samarium isotopes, using the
 auxiliary-field quantum Monte Carlo (AFMC) approach.  Before discussing these and our results, we note that there is a limit in which the quadrupole shape parameters are sharply defined; this limit corresponds to Elliott's $SU(3)$ model~\cite{su3}, when the variances of $\langle Q^2 \rangle$ and  $\langle Q^3 \rangle$ are strictly zero, as are the variances of $\beta$ and $\gamma$. This $SU(3)$ limit can be used to test our calculation of the variances, which indeed vanish up to numerical accuracy.  Shell-model calculations of  $\frac{\sigma(\hat Q^2)}{\langle \hat Q^2 \rangle}$ for the ground state and the first excited 0$^+$ of $^{42}$Ca were reported in Ref.~\cite{kasia:18}, with values of 1.96 and 0.40, respectively. It is thus obvious that it is not meaningful to specify a value for the $\beta$ parameter in the ground state. In contrast, the effective value of $\beta$ for the excited state has only a 20\% uncertainty, and we can interpret this state as a $\beta$-soft state. 
 In the AFMC calculation of Ref.~\cite{yoram:17},  values of  $\frac{\sigma(\hat Q^2)}{\langle \hat Q^2 \rangle}$ ranging from 0.57 to 0.27 were obtained for $\beta$ values ranging from 0.106  to 0.230. The effective value of $\beta$ and its variance were found to be anti-correlated, with larger values of $\beta$ associated with smaller fluctuations. 

 \begin{table}[h]
\caption{Effective deformation parameters $\beta $, $\gamma$ and their fluctuations for the nuclei discussed in the text.  } 
\label{tab1}
    \begin{tabular*}{\linewidth}{@{\extracolsep{\fill}}lcccccccr}
\hline \noalign{\smallskip}
  &  $\beta$
 &  $\Delta \beta$
 &    $ \frac{\sigma{\langle \hat Q^2 \rangle}}{\langle \hat Q^2 \rangle}$ 
  &   $\gamma$ 
 &    $ \frac{\sigma{\langle \hat Q^3 \rangle}}{\langle \hat Q^3 \rangle}$
  &  $\delta_{cv}$
   &     $  \sigma(\cos 3\gamma)$ 
 &      $\gamma$ range \\   
 \noalign{\smallskip}\hline\noalign{\smallskip}
$^{20}$Ne  &   0.62 &  0.07   & 0.24   &  3$^{\circ}$     &  0.44  &  0.10 & 0.09  &   0$^{\circ}$   ---  9$^{\circ}$ \\
\hline \noalign{\smallskip}
$^{22}$Ne  &   0.59  &  0.08  & 0.27    & 13$^{\circ}$    &  0.53  &  0.12 &  0.21  &   2$^{\circ}$   ---  18$^{\circ}$    \\
\hline \noalign{\smallskip}
$^{24}$Mg    & 0.60  &  0.07  & 0.25   & 18$^{\circ}$   &  0.42  & 0.07 & 0.21  &  12$^{\circ}$   ---  22$^{\circ}$     \\
\hline \noalign{\smallskip}
$^{28}$Si    &   0.46  &  0.09 & 0.41     & 50$^{\circ}$   &  0.91 &   0.32 & 0.42   &  38$^{\circ}$   ---  60$^{\circ}$    \\
\hline \noalign{\smallskip}
$^{48}$Cr  &   0.31 & 0.06  & 0.41& 13$^{\circ}$  & 0.84  &   0.32 & 0.30 & 0$^{\circ}$   ---  20$^{\circ}$  \\ 
\hline \noalign{\smallskip}
$^{34}$Si     & & & & & & & & \\
     0$^+_{1}$                  & 0.18 & 0.10  & 1.07   &   40$^{\circ}$  & 3.61  &  1.72 &   1.52 &  0$^{\circ}$   ---  60$^{\circ}$  \\ 
      0$^+_{2}$                 & 0.42 & 0.08 &  0.37 &   40$^{\circ}$  & 1.14  &  0.17 & 0.53 & 30$^{\circ}$   ---  60$^{\circ}$\\ 
\noalign{\smallskip}
\hline \noalign{\smallskip}
$^{44}$S  & & & & & & & & \\
    0$^+_{1}$  & 0.30 & 0.07 & 0.43  &   27$^{\circ}$  & 5.46  &  0.46 &   0.87 &  0$^{\circ}$   ---  45$^{\circ}$ \\ 
  0$^+_{2}$  &  0.31 & 0.07 &  0.43 &   21$^{\circ}$  & 1.63  &   0.36 & 0.67 & 0$^{\circ}$   ---  34$^{\circ}$ \\ 
\noalign{\smallskip}  
\hline \noalign{\smallskip}
\hline \noalign{\smallskip}
$^{68}$Ni   & & & & & & & & \\
0$^{+}_{1}$  & 0.12 & 0.07 & 1.10 &   37$^{\circ}$ & 5.10& 1.85 &1.66 & 0$^{\circ}$  -- 60$^{\circ}$\\ 
  0$^{+}_{2}$ & 0.20 & 0.07 & 0.65  &  36$^{\circ}$ & 3.43 &  0.24 & 1.12 &   12$^{\circ}$  --  60$^{\circ}$  \\
  0$^{+}_{3}$ & 0.29 & 0.05 & 0.29  & 14$^{\circ}$  & 0.62 & 0.13 &  0.26 &  0$^{\circ}$   -- 20$^{\circ}$\\ 
 \hline \noalign{\smallskip} 
$^{64}$Cr   & 0.29 & 0.06 & 0.35 & 16$^{\circ}$  & 0.84 & 0.24 & 0.35  &  0$^{\circ}$   --  24$^{\circ}$\\ 
\hline \noalign{\smallskip}
$^{70}$Zn  & 0.23 & 0.06 & 0.50 & 29$^{\circ}$  & 13.57 & 2.45 & 0.77  &  12$^{\circ}$   --  45$^{\circ}$\\ 
\hline \noalign{\smallskip}
$^{76}$Ge   & 0.25 & 0.03 & 0.27 & 28$^{\circ}$  & 3.59 & 1.0 & 0.45 &  18$^{\circ}$   --  36$^{\circ}$\\ 
\hline
\end{tabular*}
\end{table}

In the upper-right part of Table~\ref{tab1} we show our results for the parameter $\beta$ and its fluctuations for several nuclei in the $sd$ shell with the interaction USD-A~\cite{usdb}, in the $pf$ shell with KB3~\cite{kb3}, and in the $sd-pf$ shell with the interactions SDPFU-MIX \cite{sdpfu-mix} and SDPFU \cite{sdpfu}. 
In all cases we compute the mass quadrupole 
with a Dufour-Zuker isoscalar effective factor of 1.77~\cite{duzu}. The results pertaining to the charge distribution are very similar.  In almost all cases we find values  of  $\frac{\Delta \beta}{\beta}$ around 20\% or less, and
 the reference to an intrinsic structure is justified. 
 
Two rows of Table~\ref{tab1} are devoted to $^{34}$Si. This neutron-rich isotope  
 is doubly magic~\cite{34Si}, and its first excited state is a 0$^+$~\cite{rotaru} deformed state. The values shown in Table~\ref{tab1} confirm that
 the ground state is indeed spherical, and with a value of  $\frac{\sigma \langle Q^2 \rangle}{ \langle Q^2 \rangle}=1.07$, the notion of a well-defined intrinsic state is meaningless. In contrast, the excited  0$^+$ state can clearly be interpreted as  deformed.  The isotope $^{44}$S belongs to the $N=28$ island of inversion (IoI) and its collective character has been much debated~\cite{gaude}. Here we find that the ground state and the first excited 0$^+$ state in $^{44}$S are both deformed with similar effective values of  $\beta$.
  
\section{Fluctuations of $\gamma$}
Next we calculate and interpret the uncertainties in the $\gamma$  angle which arise from the non-vanishing variances of ${\langle Q^2 \rangle}$  and  ${\langle Q^3 \rangle}$.   In particular, we compute  the variance of $\cos 3\gamma$ using
\begin{equation}\label{sigma-cos3g}
\frac{\sigma^2(\cos 3\gamma)}{(\overline{\cos 3{\gamma}})^2} = \frac{\sigma^2{\langle \hat Q^3 \rangle}}{\langle \hat Q^3 \rangle^2} + \frac{9}{4}  \frac{\sigma^2{\langle \hat Q^2 \rangle}}{\langle \hat Q^2 \rangle^2}  -  3
 \frac{\langle \hat Q^5 \rangle - \langle \hat Q^3 \rangle \langle \hat Q^2 \rangle}{\langle \hat Q^3 \rangle \langle \hat Q^2 \rangle}  \;.
\end{equation}
Eq.~(\ref{sigma-cos3g}) requires the knowledge of all the moments up to the sixth order, including the fifth-order invariant that appears in the covariance term of $\hat Q^2$ and $\hat Q^3$, which we denote by $-3\delta_{cv}$.
Eq.~(\ref{sigma-cos3g}) becomes singular for $\gamma=30^\circ$ (i.e., $\langle Q^3 \rangle$=0), in which case the relevant formula is
\begin{equation}
\sigma(\cos 3\gamma) = \sqrt{\frac{7}{2}}  \; \;  \frac{\langle \hat Q^6 \rangle^{1/2}}{\langle \hat Q^2 \rangle^{3/2}} \;.
\end{equation}

%

 In upper-left part of Table~\ref{tab1},  we list the results for the effective values of $\gamma$ and its fluctuations. We compute  the standard deviation 
 $\sigma(\cos 3\gamma)$ using Eq.~(\ref{sigma-cos3g}), in which the value of  $\langle \hat Q^5 \rangle$ is calculated from Eq.~(\ref{Q5}). 
 We then compute the values of $\cos^{-1}(\cos 3\gamma \pm \sigma(\cos 3\gamma))$ 
 and  determine the corresponding range 
 of the values of $\gamma$.  We often  find values which are outside the allowed cosine range, in which case we set $\gamma=0^\circ$ or $\gamma= 60^\circ$. The ranges of $\gamma$ values in the table are surprising; in most case they span the complete prolate or oblate sectors, in which case it is at least meaningful to characterize the nucleus as prolate or oblate. Overall,  the $\gamma$ degree of freedom is so soft that it is no longer meaningful to assign to it an effective value.
 We have found only one  axial nucleus, $^{20}$Ne, and a mildly triaxial nucleus, $^{24}$Mg, in which the fluctuations of  $\gamma$  do not blur 
 its effective value. Both nuclei are close to Elliott's $SU(3)$ limit.
 

\section{Shape coexistence in $^{68}$Ni revisited}

 \begin{figure*}[]
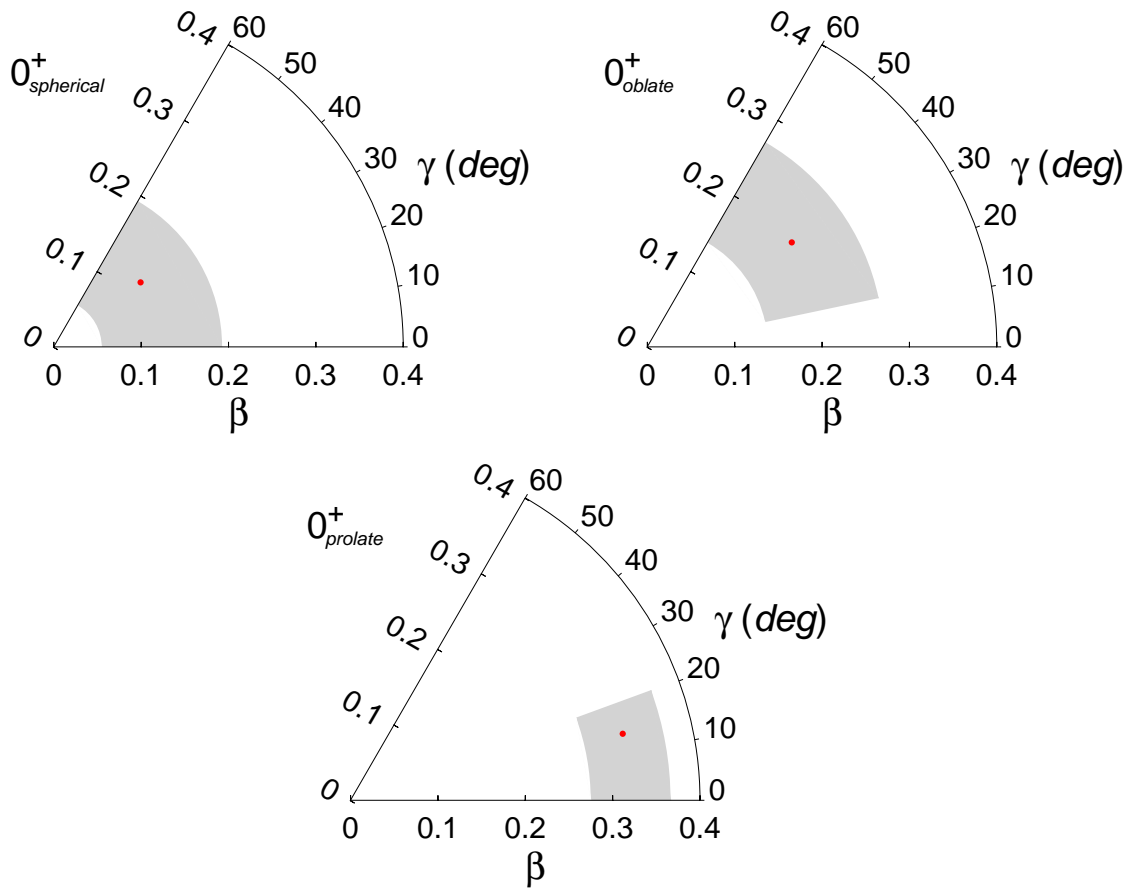

  \includegraphics[width=0.9\columnwidth]{68Ni-0+1.ps}
  \includegraphics[width=0.9\columnwidth]{68Ni-0+2.ps}
  \includegraphics[width=0.9\columnwidth]{68Ni-0+3.ps}
\caption{\label{Ni68} One sigma contours in the $\beta$-$\gamma$ sextant for the three lowest 0$^+$ states of $^{68}$Ni (K-plots). Red dots represent the effective
$\beta$,$\gamma$ values}
\end{figure*}

 With  $40$ neutrons and $28$ protons, $^{68}$Ni is a candidate to doubly magic nucleus far from stability that displays shape coexistence. It has a $0^+$ ground state that is often said to be spherical, and two excited 0$^{+}$ states, one that is considered weakly deformed oblate and another that is a strongly deformed prolate. These states have been the subject of several recent experimental studies~\cite{sorlin2002,dijon,recchia}.  
We explore the nature of these three 0$^{+}$ states using the quadrupole invariants with CI shell model  
 wavefunctions computed with the LNPS valence space and effective interaction~\cite{lnps}. Our results are summarized in 
 Table \ref{tab1} and Fig.~\ref{Ni68}. We refer to the plots shown in this figure as K-plots.
  With $\sigma{\langle Q^2 \rangle} \sim  \langle Q^2 \rangle$  and  $\gamma $ completely undefined, the notion of a 'spherical' shape is not meaningful for the ground state of  $^{68}$Ni. 
  We find the first excited 0$^{+}$  to be very soft in $\beta$, but we can still call it deformed.
   One could say that it is oblate, although its respective $\gamma$ values cover a large
   fraction of the prolate sector.  The second excited 0$^{+}$ has a larger and relatively well-defined value of $\beta$. The spread
   in  $\gamma$ is very large (20$^{\circ}$), but at least it remains confined into the prolate sector.   
  
   We also include in Table \ref{tab1} results for $^{64}$Cr, a nucleus at the centre of the $N=40$ IoI, to
   confirm that they are nearly identical to those of the well-deformed excited $0^+$ band-head of $^{68}$Ni.
    This provides another evidence of the role of shape coexistence
    in doubly magic nuclei as the portal to the adjacent IoI. 
    
\section{Rigid triaxiallity in $^{76}$Ge and $^{76}$Se falsified}  

 Next, we discuss two nuclei that are relevant to neutrinoless double beta decay searches, $^{76}$Se and $^{76}$Ge,  whose effective $\gamma$ values  are very close to  $30^{\circ}$.  $^{76}$Ge was described in Ref.~\cite{ge76} as an example of a rigid triaxial nucleus.  However, our calculations with the LNPS interaction and standard effective charges, which reproduce very well its spectroscopic properties,  show a rather large spread ($\sim 18^{\circ}$) in $\gamma$, not fully compatible with rigid triaxiality.  This spread we find in $\gamma$ also disagrees with the one obtained in a recent analysis~\cite{ge76a}.   Kumar's invariants were extracted from
 experimental data and from CI shell model calculations in another recent article entitled ``Trixiality in $^{76}$Se''~\cite{se76}. 
 Using the same interactions and valence spaces used in the latter article, we find the
 dispersion in $\langle \hat Q^2 \rangle$ to be 27\% and 36\% (for each of the two interactions used), and the dispersion in $\gamma$ to be $36^{\circ}$  and $35^{\circ}$, respectively. These values are much larger than those concluded in Ref.~\cite{se76}, demonstrating the challenge of calculating the expectation values of higher-order invariants in the CI shell model.

\section{Are doubly magic nuclei spherical?}  As already discussed in the context of the `spherical' $^{68}$Ni ground state, the notion of a well defined shape for nuclei usually called spherical, can be meaningless when  inferred from shell-model Kumar's invariants.   We have examined two classical doubly magic nuclei,  $^{48}$Ca and $^{56}$Ni, in the full $pf$ shell using the KB3G interaction.  Kumar's invariants lead to  $\beta$=0.15, $\gamma$=33$^{\circ}$  for $^{48}$Ca,  and  $\beta$=0.21, $\gamma$=41$^{\circ}$  for $^{56}$Ni. In both cases,  $\sigma{\langle Q^2 \rangle} \sim  \langle Q^2 \rangle$  and  the spread of $\gamma$ at 1$\sigma$ $\sim$ 60$^{\circ}$. Hence, their shapes are completely undefined, making the expression `spherical shape' a quantal oxymoron. 
                                  
\section{Conclusion} We introduce a novel method for calculating higher-order quadruple invariants in the CI shell-model approach, enabling accurate calculation of the fluctuations in both intrinsic shape parameters $\beta$ and $\gamma$.  
In the mass regions we explored, we find,  with the exception of 
$^{20}$Ne and  $^{24}$Mg, that $\gamma$ is characterized by large fluctuations.  In particular, we find that the $A=76$ isobars of germanium and selenium exhibit significant $\gamma$ fluctuations in their ground states, contrary to recent claims that these nuclei are rigid triaxial.  In general, it is more meaningful to describe the nuclear state in terms of a probability distribution of intrinsic shapes~\cite{yoram:18} rather than assigning particular values to $\beta$ and $\gamma$. 
 Our analysis has direct consequences for the interpretation of laboratory-frame results in the $(\beta, \gamma)$
 plane of intrinsic shapes, as is often done in mean-field (and beyond) approaches~\cite{bmf}.
 We find that it is particularly important to estimate the dispersion in $\gamma$ before assigning to it a well-defined value. 
  \acknowledgements
  
 We thank Tom\'as Rodr\'{i}guez-Frutos for valuable discussions.
 This work was supported in part by the Ministerio de Ciencia, Innovaci\'on y Universidades (Spain),  Severo Ochoa Programme SEV-2016-0597
 and grant PGC-2018-94583, and by the U.S. DOE grant No.~DE-SC0019521.


\begin{thebibliography}{80}
\bibitem{commute} Strictly speaking this holds under the assumption that the quadrupole operators $\hat Q_{2\mu}$ commute among themselves. This is not the case in the truncated shell-model space but the effect of their non-commutation is small.
\end{thebibliography}
\end{document}